\documentclass{mem}
\usepackage{natbib}
\usepackage{txfonts}
\usepackage{graphicx}
\usepackage[a4paper]{hyperref}
\idline{75}{282}
\begin{document}

\title{
Recovering the Line-Of-Sight Magnetic Field in the Chromosphere from Ca~II~IR Spectra
}

   \subtitle{}

\author{
Friedrich W{\"o}ger\inst{1} \and
Sven Wedemeyer-B{\"o}hm\inst{2} \and
Han Uitenbroek\inst{1} \and
Thomas Rimmele\inst{1}
        }

\offprints{F. W{\"o}ger}

\institute{
National Solar Observatory
PO Box 62
Sunspot, NM 88349
USA
\and
Institute of Theoretical Astrophysics,
University of Oslo,
Postboks 1029 Blindern, N-0315 Oslo,
Norway\\
\email{fwoeger@nso.edu}
}

\authorrunning{W{\"o}ger }

\titlerunning{Magentic Fields in the Chromosphere}

\abstract{
We propose a method to derive the line-of-sight magnetic flux density from measurements in the chromospheric Ca II IR line at 854.2\,nm. The method combines two well-understood techniques, the center-of-gravity and bisector methods, in a single hybrid technique. The technique is tested with magneto-static simulations of a flux tube.
We apply the method to observations with the Interferometric Bidimensional Spectrometer (IBIS) installed at the Dunn Solar Telescope of the NSO/SP to investigate the morphology of the lower chromosphere, with focus on the chromospheric counterparts to the underlying photospheric magnetic flux elements.
}

\maketitle{}

\section{Introduction}
Advances in high-resolution observation techniques have yielded data suggesting the existence of a weak-field domain below the classical canopy \citep{2006A&A...459L...9W}.  
Magnetic fields play a major role in the morphology of the chromosphere such as it is seen in the H$\alpha$ line core.
The region below referred to as ``fluctosphere'' \citep{2008IAUS..247...66W} or ``clapotisphere'' \citep{1991SoPh..134...15R} contains presumably only weak-fields.
In numerical simulations, this domain is generated by interfering (acoustic) shock waves that are excited in the photosphere and propagate into the layers above \citep[cf.][]{1994chdy.conf...47C}.

\citet{2008A&A...480..515C} have shown that the Ca II infrared triplet provides a convenient and accessible diagnostic of the chromosphere.
Confirming the shock behavior predicted by the radiation hydrodynamic numerical models in regions with no or at least weak magnetic field, it has been found that there is a strong influence of magnetic field on acoustic processes such as shock signature surpression within that regime of the solar atmosphere \citep{2001ApJ...554..424J, 2009A&A...494..269V}.
From their findings it has been concluded that this influence may be larger than generally expected, in particular because it is likely that there exist at least weak fields in this domain.

For this reason, knowledge of the chromospheric magnetic field topology and dynamics is of significant importance for the understanding of the solar atmosphere.
Yet, these properties are poorly known due to limitations set by current observational techniques, in particular when gathering data of weak-field regions in the quiet Sun.
In addition, the interpretation of chromospheric spectro-polarimetric data has proven to be difficult as many complicating effects have to be taken into account such as NLTE conditions.

Here we present and interpret high spatially resolved spectro-polarimetric measurements in the fluctosphere above strong photospheric magnetic features.

\section{Model}
A disadvantage of the robust center of gravity (COG) method \citep[e.g.][]{2003ApJ...592.1225U} to derive LOS magnetic flux is the lack of height resolution.
To recover the height distribution of the magnetic flux, a bisector analysis is often employed but it is susceptible to noise.
Thus, to enhance the reliability of a topology analysis, we suggest a technique that combines bisector and COG method.

\subsection{The Hybrid-COG method}
The following steps are performed in our algorithm:
\begin{enumerate}
\item For both, the \mbox{I+V} and \mbox{I-V} spectra, find the two points ($\lambda_{a}$ and $\lambda_{b}$) of intersection of the line profile with a horizontal line (the points at which the red and blue line wing have the same intensity). 
\item Compute the COG of the I+V and I-V spectra (using the two acquired wavelength ranges) with
\begin{equation}
\lambda_{+} = \frac
{\int_{\lambda_{a}}^{\lambda_{b}}\mbox{d}\lambda\; \lambda\; \{\mbox{(I+V)}(\lambda_{a}) - \mbox{(I+V)}(\lambda)\}}
{\int_{\lambda_{a}}^{\lambda_{b}}\mbox{d}\lambda\; \{\mbox{(I+V)}(\lambda_{a}) - \mbox{(I+V)}(\lambda)\}}
\end{equation}
for \mbox{I+V} and analogously $\lambda_{-}$ for \mbox{I-V}.
\item Compute the value for the magnetic flux density along the line-of-sight: calculate the difference of the COG positions of the \mbox{I+V} and \mbox{I-V} profiles, using
\begin{equation}
\mbox{B}_{\mbox{\tiny LOS}} = \frac{\lambda_{+}-\lambda_{-}}{2} \  \frac{4\ \pi\ m\ c}{e\ g_{L}\ \lambda_{0}^{2}},
\end{equation}
where $\lambda_{0}$ is the central wavelength of the line, $\lambda_{\pm}$ are the wavelength positions of the centroids of I$\pm$V, $g_{L}$ is the effective Land{\'e} factor, and $e$ and $m$ are the electron charge and mass, respectively, in SI units \citep[see][and references therein]{2003ApJ...592.1225U}.
\end{enumerate}
Step 1 allows to derive the magnetic flux density at different formation height ranges, as it is done in the genuine bisector approach.
The even cuts are necessary to avoid bias in the measurement of the COG value.
Obviously, the technique will work best for spectral lines without an emission peak, as it would make Step~1 (and Step~2) difficult by creating four instead of two points of intersection and also bias the computation of the COG.
Asymmetries in the Stokes V profiles, such as those found e.g. in a detailed analysis of a Quiet Sun region near network by \citet{2007ApJ...670..885P} on a common basis, pose a further problem.
They can lead to biased measurements of magnetic flux.
To test the proposed method, we have applied it to synthetic profiles that inhibit similar behavior.

\subsection{Check of method}
\begin{figure*}
\centering
\includegraphics[width=\textwidth]{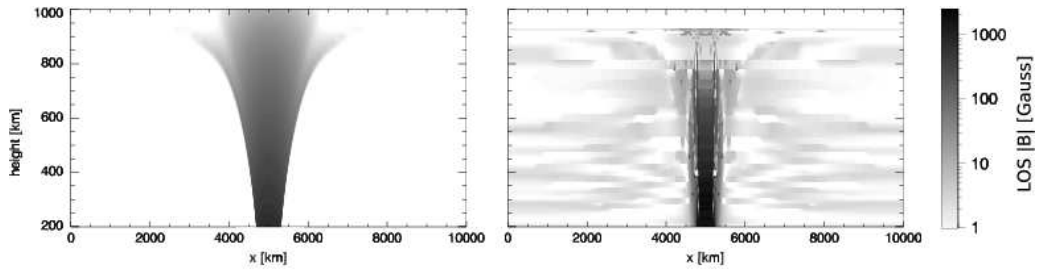}
\caption{Result of the accuracy test of the proposed hybrid bisector-COG method. a) input model, and b) recovered flux density. The height scale in b) was derived using calculations of $\tau_{\lambda}=1$. The gray scale encodes the magnetic flux density and is the same for both panels.}
\label{fig:methodtest}
\end{figure*}

A magneto hydrostatic model of a funnel-like expanding flux tube served as test case for the feasibility of our approach.
The foot point of the flux tube in the deep photosphere has a diameter of 500\,km and a magnetic field strength of 2500\,Gauss.
For the Ca~II~IR line, synthetic NLTE spectra for all four Stokes parameters were computed.
These served as input for the hybrid-COG method.
The resulting output is compared to the initial flux tube model (Fig.~\ref{fig:methodtest}).

The topology of the flux tube in photosphere and chromosphere as well as the absolute magnetic flux density was recovered properly, and adjacent intensity levels show no significant jumps in the result thus exhibiting little noise.
This is likely due to the inclusion of many points in the computation of the COG as opposed to the genuine bisector method.
However, the three-dimensional topology in the higher layers cannot be directly interpreted because a bisector does not necessarily correspond to a thin atmospheric layer.

To be able to compare our results with the model input, we computed the effective formation height for the Ca II infrared line by averaging the geometrical heights of the locations where $\tau_{\lambda}=1$ over each of the wavelength intervals computed in Step~1.

The line formation height decreases from line core towards line wing and continuum.
Furthermore, the formation height is affected by the presence of magnetic fields: due to the ``Wilson effect'' $\tau_{\lambda}=1$ is different in magnetic and non-magnetic regions.
As an example, compared to the ambient plasma the opacity within the model flux tube is lowered by the strong magnetic field.
This leads to the formation of $\tau_{\lambda}=1$ at a height that is about 200\,km lower in the flux tube than in the non-magnetic atmosphere, which has been compensated for in Fig.~\ref{fig:methodtest}~b).

Comparing both panels of Fig.~\ref{fig:methodtest} allows to test the reliability and estimate the error of the proposed technique.
Line properties such as formation height, and input parameters to technique such as the wavelength interval for Step~1 limit the recoverable height interval.
The height distribution of the magnetic field of the modeled flux tube was recovered qualitatively.
However, regions below the ``canopy'' of the flux tube show a weak signal in the reconstructed topology with strengths of up to $(6 \pm 1)$\,G.
Because of discontinuities in the magnetic field along the LOS, the Stokes V spectra show spikes and thus bias the I+V and I-V spectra.
This leads to a fake magnetic signal in the reconstructed topology.

\section{Data}
In this section, we present observations of a quiet Sun region located at disk-center using the Ca\,II infrared line at 854.2\,nm on September 22, 2008.
Several persistent magnetic bright points are visible in the field of view, forming a network element (Fig.~\ref{fig:3dtop}~a)--b)).

In dual-beam spectro-polarimetric mode, the IBIS narrowband channel was configured to scan the Ca II IR line at 854.2\,nm using six modulation states with exposures of 90\,ms.
An overall cadence of about 1 minute was achieved by sampling the line with 30 wavelength step that were separated by 4.3\,pm, where the full width at half maximum (FWHM) transmission of IBIS at this wavelength is 4.6\,pm.
The output data was a data cube with two spatial and one spectral dimension.
Unfortunately, even after averaging 3 full scans after data reduction, the signal level of the Stokes Q and U component was below the noise level -- no linear polarization was measured.

To gather context data about the magnetic field in the photosphere, a spectral region that includes both the Fe lines at 630.1 and 630.2 as well as the telluric blend separating these two lines was observed with a temporal offset of 1 minute.
We have scanned the line with 30 wavelength steps with a stepwidth of 3.2\,pm, where the FWHM of IBIS at this wavelength is 2.3\,pm.

\subsection{Data Calibration}

For the broadband images, the standard calibration method of dark- and gain-table correction was applied.
In a subsequent step, the images were reconstructed with a speckle interferometric code adapted for use with high-order adaptive optics corrected data \citep{2008A&A...488..375W}.
The reconstructed images were used in a later step to reduce atmospheric differential image motion in both IBIS' broad- and narrowband channels.

As detailed in \citet{2006SoPh..236..415C}, IBIS' narrowband channel consists of two Fabry-Perot interferometers, which are located in a collimated beam.
This optical setup produces data that requires an elaborate procedure to take into account the blueshift in each pixel of a single narrowband exposure \citep{2006SoPh..236..415C}.
This procedure has been detailed by \citet{2006A&A...450..365J}.

The calibration of the polarimetric data is accomplished with procedures originating from those of the Advanced Stokes Polarimeter (ASP) \citep[e.g.][]{1993ApJ...418..928L} adapted for the IBIS instrument.
Residual crosstalk between the Stokes parameters has been removed manually.
The quality of IBIS spectro-polarimetric data has recently been compared with that of the spectrograph (SOT/SP) on-board the HINODE satellite
The result is that both instruments deliver very similar spectro-polarimetric data, which indicates accurate calibration \citep{judge}.

The adapted procedure to calibrate IBIS' spectroscopic and spectro-polarimetric data will be described in detail in a forthcoming publication \citep{tritschler}.

\subsection{Application of Hybrid-COG}
Analysis of our data set with the suggested hybrid COG method has been accomplished using 101 intensity values ranging from 18.5\% (Ca II IR line core) to 32\% of the local continuum intensity.
To reduce salt and pepper noise introduced by the measured spectra, a median filter with 5$\times$5 pixels (corresponding to 0.85$\times$0.85\,arcs$^2$) is applied to the data.

\begin{figure*}
\centering
\includegraphics[width=\textwidth]{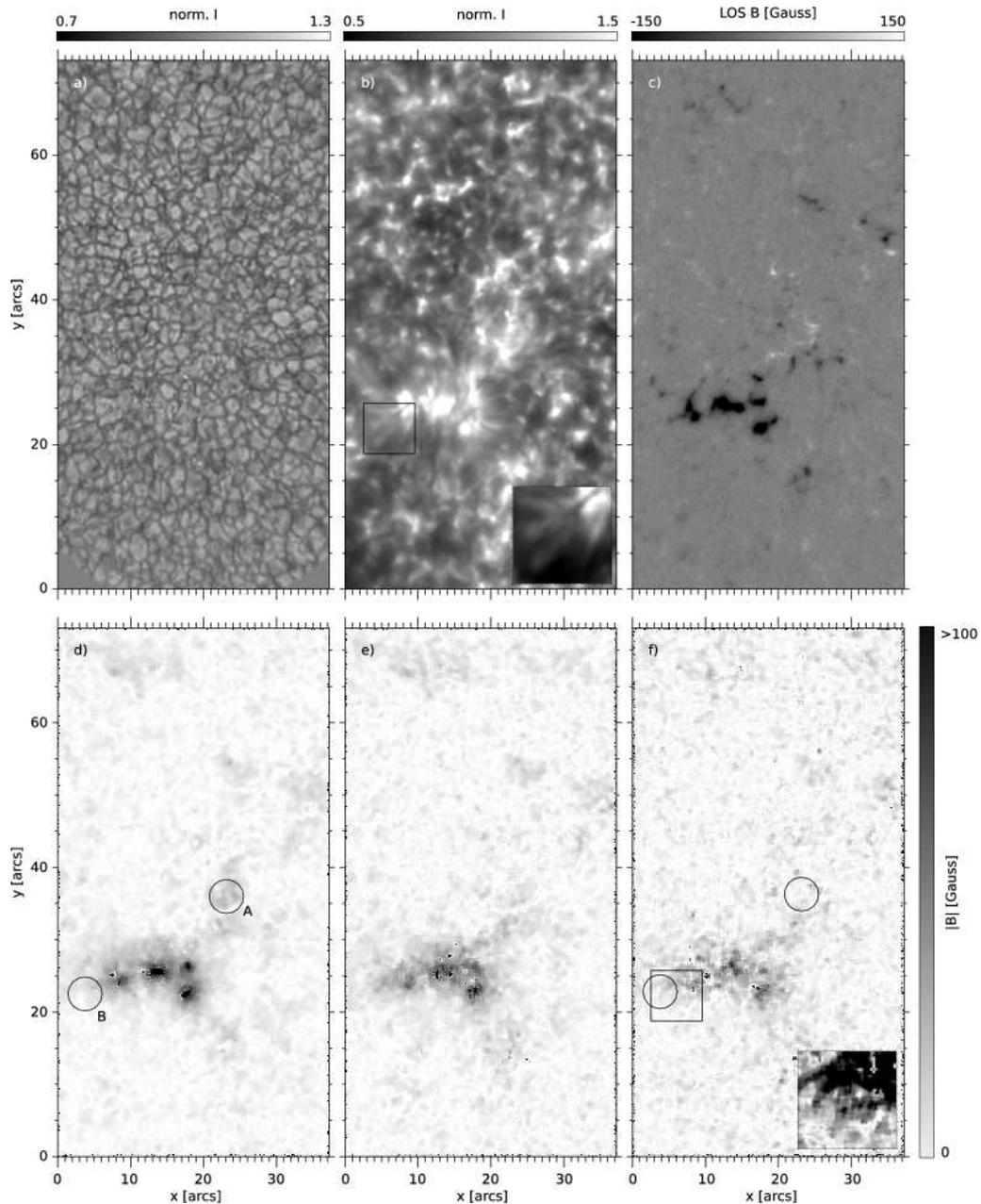}
\caption{a) Continuum intensity around 852\,nm, b) Ca II IR (854.2\,nm) line core intensity, c) LOS magnetogram computed from Fe I (630.2\,nm).
Cuts through the recovered 3D topology of the magnetic flux density by applying the hybrid bisector-COG method to the Ca II IR line scan.
Panel d) is at 32\% of the local continuum intensity (photospheric), e) at 20.2\% and f) at 18.5\% (chromospheric/``fluctospheric'').
The subpanel in f) shows a contrast enhanced close-up of the fibrils marked by the box.
At location~A, a patch of magnetic flux along the line-of-sight seems to disappear with increasing height (also see subpanel).
Location~B is an example for the appearance of a filament like structure with height.
}
\label{fig:3dtop}
\end{figure*}
The COG method was applied to the Fe\,I line data to gather information about the underlying photospheric magnetic flux density which amounted to about 350\,Gauss within the strongest feature.
The small-scale magnetic elements were likely not resolved.
The Fe\,I line is formed within a relatively thin atmospheric layer on the Sun, rendering the COG method sufficient, whereas the Ca\,II IR line has contributions from layers covering 1000\,km.

The thus computed map of magnetic flux density is also a check for the values retrieved from the COG method using the Ca II IR line, which delivers a value of 250\,Gauss using a cut at $\sim$50\% of continuum intensity.
As expected, the formation height at that level appears to be higher than that of the Fe\,I line.

Using the suggested hybrid COG method, starting at a level of 32\% of the local continuum intensity, this value drops to 125\,Gauss.
This result is not surprising because a restriction of the method to the line core is equivalent to a restriction to higher layers.
Further results of our hybrid bisector-COG analysis can be viewed in Fig.~\ref{fig:3dtop}~d)--f).

Two situations are of interest in the field of view.
At location A in Fig.~\ref{fig:3dtop}~d), a patch of likely unresolved LOS magnetic flux disappears as when including only the few points in the line core in the hybrid COG method, and thus restricting the analysis to the highest layers.
It appears as if the magnetic flux does not reach the corresponding heights.
On the other hand, at the location marked B in Fig.~\ref{fig:3dtop}~d), a diffuse patch starts to show a filament structure at the high layers.

Overall, it appears that (i) the strong photospheric magnetic structure in the FOV fragments, and (ii) the flux density along the LOS, which corresponds to the vertical component of the field (the data were observed at disk-center of the Sun), becomes smaller and weaker.
A magnetic funnel expanding with height would show similar characteristics: the magnetic field becomes increasingly horizontal with height, and the vertical flux component decreases as the the horizontal component, that is unfortunately currently not measurable with the IBIS instrument, increases.

\section{Conclusions}
The focus of this contribution is the derivation of the absolute LOS magnetic flux from spectro-polarimetric observations of the Fe\,I (630.2\,nm) and Ca\,II\,IR (854.2\,nm) lines.

We suggest to compute the 3D topology of the absolute LOS magnetic field flux per pixel using a new technique that is based on understood methods.
The technique has been analyzed for accuracy for observations in a Ca\,II\,IR line in a Quiet Sun region near disk center.
We were capable to recover the topology of a strong magnetic structure extending from photosphere to low chromosphere, and find that the magnetic flux becomes weaker with height developing a filament like structure as its height increases.
The suggested method is not capable to infer filling factors.
In an unresolved magnetic element, the field strength thus cannot be recovered.

While our findings suggest the existence of a funnel-like structure in the chromosphere, the location of the horizontal magnetic field remains unclear.
Overall, we believe that a static flux tube funnel is a model too simplified to reflect realistically the conditions present in the chromosphere.

Understanding the chromospheric energy balance requires spatially highly resolved measurements of the magnetic field in the chromosphere.
Such data, in spite of advanced instrumentation available at modern telescopes today, still lack the signal-to-noise ratio to gain the accuracy needed to detect weak fields at the temporal resolution which is implied by the variations seen in chromospheric diagnostics.
The only way to address this problem is building a new generation of telescopes with larger apertures.

Interpretation of weak magnetic field measurements in the chromosphere is at least of equal difficulty as their gathering.
New tools need to be developed to better understand the chromosphere as an important layer between the photosphere and corona; for example, the height origins of the Stokes signals are unclear and do not only depend on contribution function but also on height distribution of the magnetic field strength.
Discontinuities and steep gradients in magnetic field strength remain to be fundamental problems: they are difficult to detect and to recover from measurements with methods available today.

\begin{acknowledgements}
SWB acknowledges support through a Marie Curie Intra-European Fellowship of the
European Commission (6th Framework Programme, FP6-2005-Mobility-5, Proposal No.
042049).
The authors would like to thank A. Tritschler for support during the data reduction.
\end{acknowledgements}


\bibliographystyle{aa}

\end{document}